\documentclass[a4paper,11pt]{article}

\usepackage{contribution}
\usepackage{overpic}

\newcommand{\weblink}[2][]{%
    \ifthenelse{\equal{#1}{}}%
    {\textnormal{\url{#2}}}%
    {\textnormal{\href{#2}{#1}}}%
}

\newcommand{\acknowledgements}[1]{%
  \bigskip\bigskip
  \textsf{\textbf{\Large Acknowledgements}} \\[2ex]
  {#1}
  \bigskip
}

\def\beq{\begin{equation}}
\def\eeq#1{\label{#1}\end{equation}}
\def\eeqn{\end{equation}}

\def\beqa{\begin{eqnarray}}
\def\eeqa#1{\label{#1}\end{eqnarray}}
\def\eeqan{\end{eqnarray}}

\let\bar=\overbar

\def\etal{{\it et al.}}

\def\Dslash{\not{\hbox{\kern-4pt $D$}}}
\def\dslash{\not{\hbox{\kern-2pt $\del$}}}

\def\msb{{\bar{\ssstyle M \kern -1pt S}}}

\newcommand{\contribution}[7][]{%
  \clearpage
  \thispagestyle{plain}
  \ifthenelse{\equal{#1}{}}
  {\hypersetup{pdftitle={#2}}}
  {\hypersetup{pdftitle={#1}}}
  \hypersetup{pdfauthor={{#3} {#4}}}
  {\centering\normalfont\LARGE\bfseries\sffamily #2 \par\nobreak}
  \lhead{}
  \chead{%
    \textit{\footnotesize XIV International Conference on Hadron Spectroscopy
      (\weblink[\textit{hadron2011}]{http://www.hadron2011.de}), 13-17 June 2011, Munich, Germany}%
  }
  \rhead{}
  \bigskip
  \begin{center}
    {#3} {#4}\ifthenelse{\equal{#6}{}}{}{\footnote{\weblink[#6]{mailto:#6}}}
    \ifthenelse{\equal{#7}{}}{}{#7} \\
    \textit{#5}
  \end{center}
  \bigskip
}

\renewcommand{\abstract}[1]{%
  \begin{center}
    \begin{minipage}{0.85\textwidth}
      \begin{footnotesize}
        #1
      \end{footnotesize}
    \end{minipage}
  \end{center}
  \bigskip
}

\begin{document}

%
%
%
%
%
{  

\contribution[Measurement of the double polarisation observable E] 
{Measurement of the double polarisation observable E in the reactions $\overrightarrow{\gamma}\overrightarrow{p}\rightarrow p\eta$ and $\overrightarrow{\gamma}\overrightarrow{p}\rightarrow p\pi^{0}$} 
{Jonas}{M\"{u}ller}  
{Helmholtz-Institut f\"{u}r Strahlen- und Kernphysik, 
 Nussallee 14-16, D-53115 Bonn, Germany}  
{jonas.mueller@hiskp.uni-bonn.de}  
{and Manuela Gottschall for the CBELSA\slash TAPS Collaboration}  

\abstract{%
Polarisation observables are measured with the Crystal Barrel/TAPS experiment at ELSA for photoproduction reactions with various final states, 
using a circularly or linearly polarised photon beam and a longitudinally or transversely polarised frozen spin butanol target.
The Crystal Barrel/TAPS setup provides a nearly $4\pi$ angular coverage, a good energy resolution 
and a high detection efficiency for photons, and is therefore ideally suited to study final states comprising neutral mesons. 
Results for the measurement of the double polarisation observable E 
for the reactions $\overrightarrow{\gamma}\overrightarrow{p} \rightarrow p\pi^{0}$ and $\overrightarrow{\gamma}\overrightarrow{p} \rightarrow p\eta$ 
are presented. \\
}
%

\section{Introduction}
One of the open challenges in subnuclear physics is to understand the non-perturbative regime of QCD, including the world of the nucleon and its excitations. 
In order to extract baryon resonances in photoproduction experiments, partial wave analyses need to be performed. 
A complete experiment is required to unambiguously determine the contributing amplitudes. 
This involves the measurement of carefully chosen single and double polarisation observables.
Considering circular beam and longitudinal target polarisation, the cross section for the photoproduction of single pseudoscalar mesons can be written as
\begin{equation}
  \frac{\mathrm{d}\sigma}{\mathrm{d}\Omega} = \left(\frac{\mathrm{d}\sigma}{\mathrm{d}\Omega}\right)_0 \left(1 - p_zp_{\gamma}^{\circ}E \right).
\end{equation}
where $\left(\frac{\mathrm{d}\sigma}{\mathrm{d}\Omega}\right)_0$ is the unpolarised cross section, $p_{\gamma}^{\circ}$ the beam and $p_z$ the target polarisation\cite{barker:1975}. 

\section{Experimental setup}
The data presented has been taken with the Crystal Barrel/TAPS experiment at ELSA\cite{hillert:2006}. The setup consists of two electromagnetic calorimeters, the Crystal Barrel\cite{aker:1992} and the MiniTAPS\cite{novotny:1991} detector, covering the polar angle from $1^{\circ}$ to $156^{\circ}$ and the full azimuthal angle. Plastic scintillator detectors and a three-layer plastic fibre detector\cite{suft:2005} are used to obtain charge information. A tagged photon beam is produced from the electron beam impinging a radiator target. The Bonn Frozen Spin Target\cite{bradtke:1999} uses butanol to provide polarised nucleons.

In order to extract the double polarisation observable E, data has been taken with a longitudinally polarised target and a circularly polarised photon beam produced via helicity transfer from a longitudinally polarised electron beam at an electron beam energy of 2.35~GeV. Mean polarisation values of $p_{\mathrm{e}^-} \approx 64 \%$ and $p_z \approx 73 \%$ for beam and target were reached in the according beamtimes.

\section{Data selection and preliminary results}
The data sample was selected by choosing events with either three distinct hits in the calorimeters, two uncharged and one charged, or two uncharged hits in the calorimeters with an additional hit in a charge identification detector. The events were further selected by applying kinematic cuts on energy and momentum conservation and by cuts on the $\gamma\gamma$~invariant mass to separate the reactions $\vec{\gamma}\vec{p} \to p \pi^0$ and $\vec{\gamma}\vec{p} \to p \eta$.

\begin{figure}[bht!]
  \begin{center}
    \begin{overpic}[width=0.50\textwidth, trim=0.8cm 0.2cm 1.1cm 0.0cm, clip]{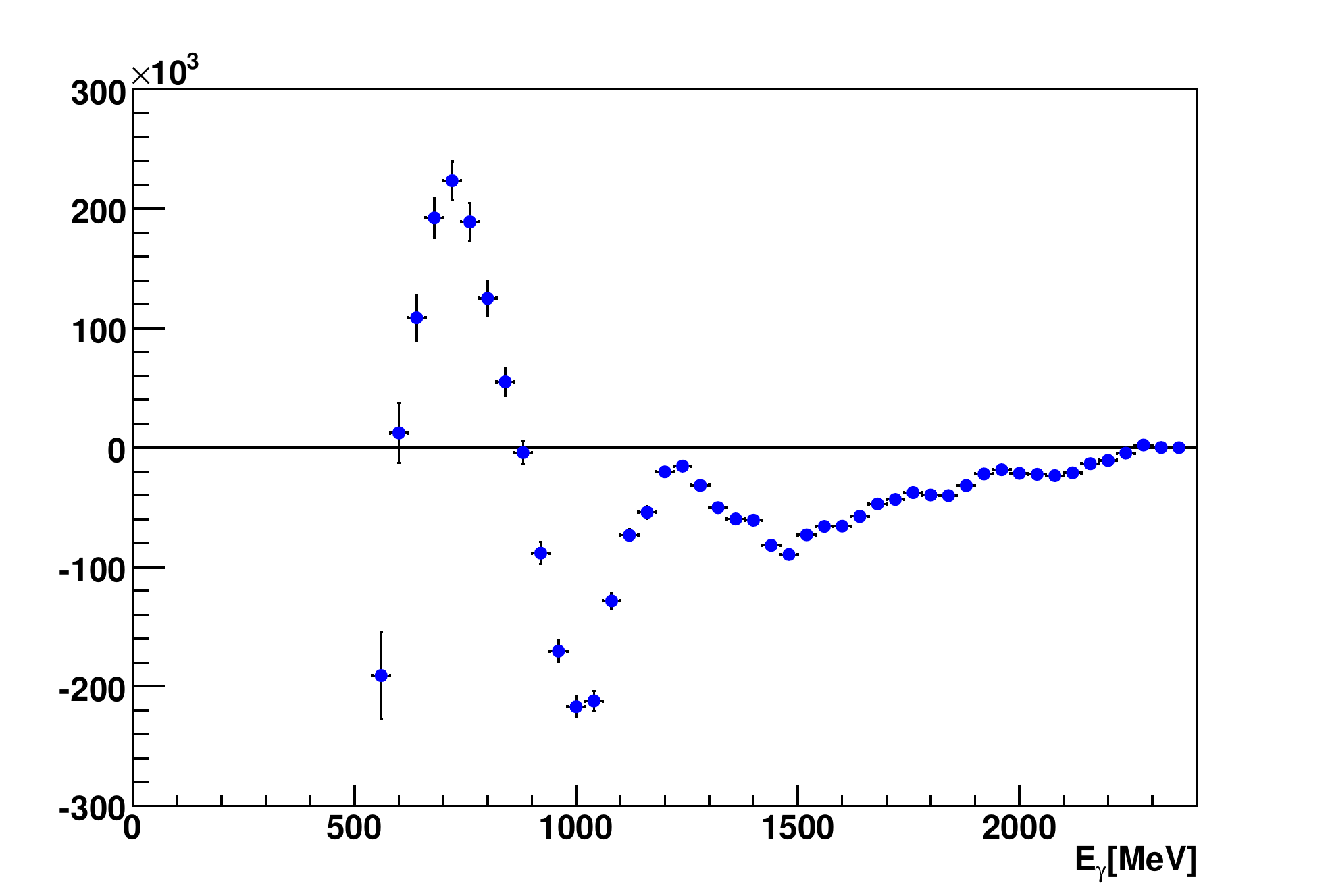}
      \put(20,11){\includegraphics[width=0.25\textwidth]{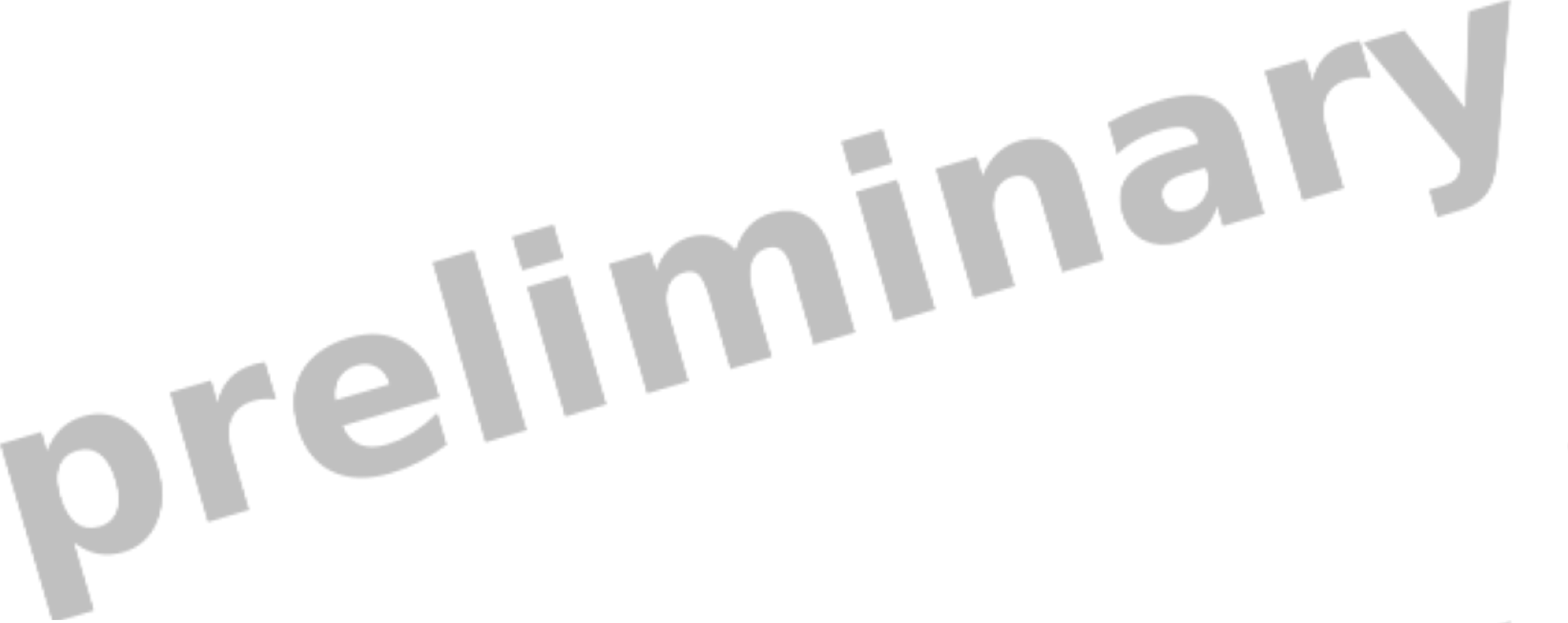}}
      \put(61.8,37.4){\includegraphics[width=0.17\textwidth]{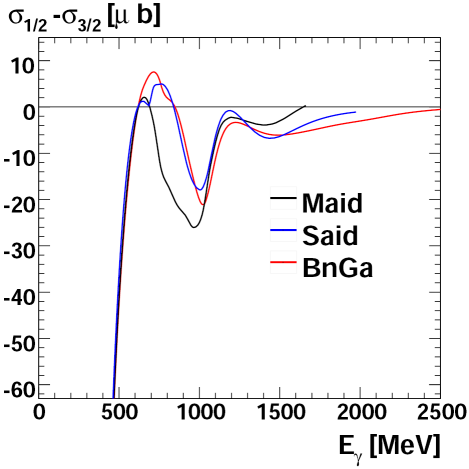}}
      \put(9.8,62.8){\begin{tiny}$N_{1/2}-N_{3/2}$\end{tiny}}
    \end{overpic}
    \begin{overpic}[width=0.49\textwidth, trim=0.2cm 0.2cm 0.0cm 0.0cm, clip]{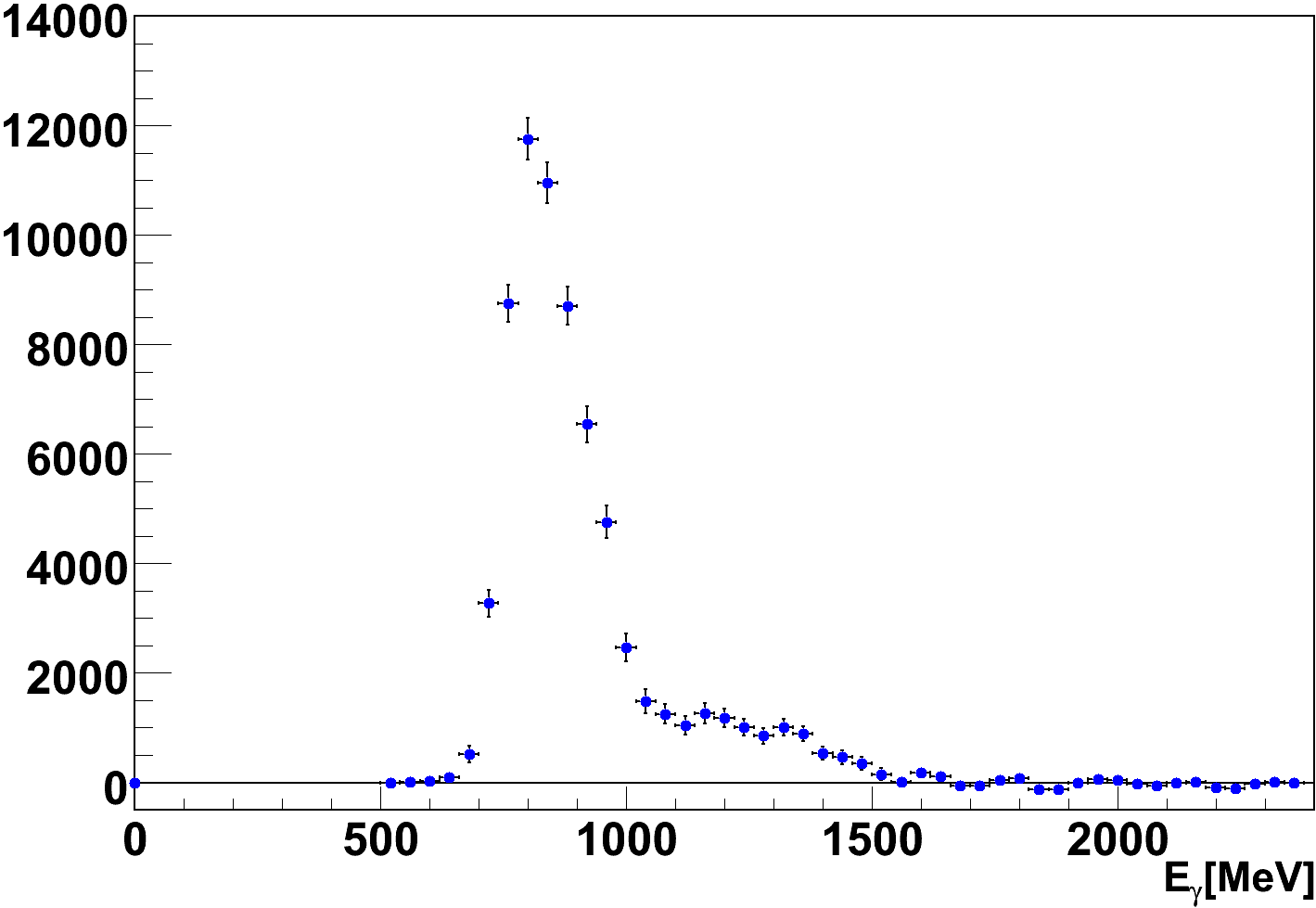}
      \put(20,11){\includegraphics[width=0.25\textwidth]{preliminary}}
      \put(58.5,34){\includegraphics[width=0.23\textwidth]{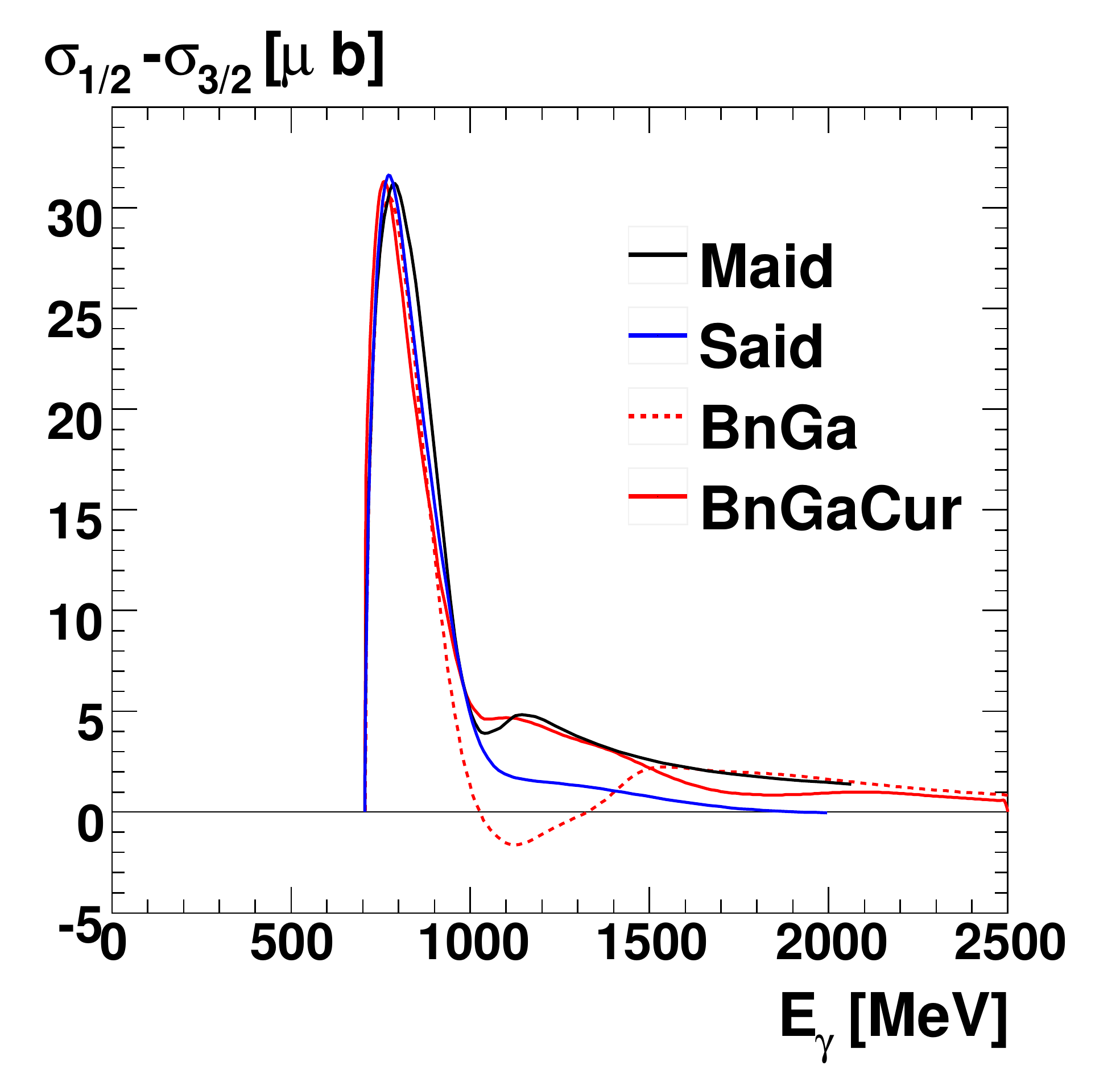}}
      \put(12,64){\begin{tiny}$N_{1/2}-N_{3/2}$\end{tiny}}
    \end{overpic}
    \caption{Polarisation weighted and acceptance corrected count rate difference for the reactions $\vec{\gamma}\vec{p} \to p \pi^0$ (left) and $\vec{\gamma}\vec{p} \to p \eta$ (right). The inlays show the predictions of the MAID\cite{maid:2007}, SAID\cite{said:2009}, and BnGa\cite{bnga}\cite{bnga2} analyses.} 
    \label{fig:cr_eta}
  \end{center}
\end{figure}

As not all protons in the butanol are polarised, two additional measurements on a carbon and a liquid hydrogen target were performed in order to determine an effective dilution factor $f_{\mathrm{dil}}$ which gives the ratio of polarisable protons to the total amount of nucleons in our data sample. 
The double polarisation observable E can then be extracted as
\begin{equation}
E = \frac{\sigma_{1/2} - \sigma_{3/2}}{\sigma_{1/2} + \sigma_{3/2}} = \frac{1}{p_{\gamma}^{\circ} \cdot p_z} \cdot \frac{1}{f_{\mathrm{dil}}} \cdot \frac{N_{1/2}-N_{3/2}}{N_{1/2}+N_{3/2}}
\end{equation}
with $\sigma_{1/2}$ and $\sigma_{3/2}$ being the cross sections and $N_{1/2}$ and $N_{3/2}$ the number of events with antiparallel and parallel spins of beam and target.

Fig. \ref{fig:cr_eta} shows the energy dependent count rate differences with antiparallel and parallel spin settings of target and beam in the reactions $\vec{\gamma}\vec{p} \to p \pi^0$ and $\vec{\gamma}\vec{p} \to p \eta$. One can clearly see the strong resonance peak due to the $S_{11}(1535)$ resonance in the $p \eta$ channel at photon energies below 1000~MeV and further to larger energies the highly debated region with possible contributions from a $P_{11}(1710)$ or a $P_{13}(1720)$. 
While the $P_{11}(1710)$ was not needed in the PWA solution named BnGa it contributes to BnGaCur, where also a reduced contribution of the $P_{13}(1720)$ was found compared to the earlier BnGa solution. The two different solutions lead to a significantly different prediction of $\sigma_{1/2}-\sigma_{3/2}$, BnGaCur showing a distribution with a higher similarity to the measured $N_{1/2} - N_{3/2}$ distribution.
In the $p \pi^0$ channel, prominent structures over the whole energy range are visible, starting with an excess of $N_{3/2}$ events due to the $P_{33}(1232)$.
\begin{figure}[htb!]
  \begin{center}
    \begin{overpic}[width=\textwidth]{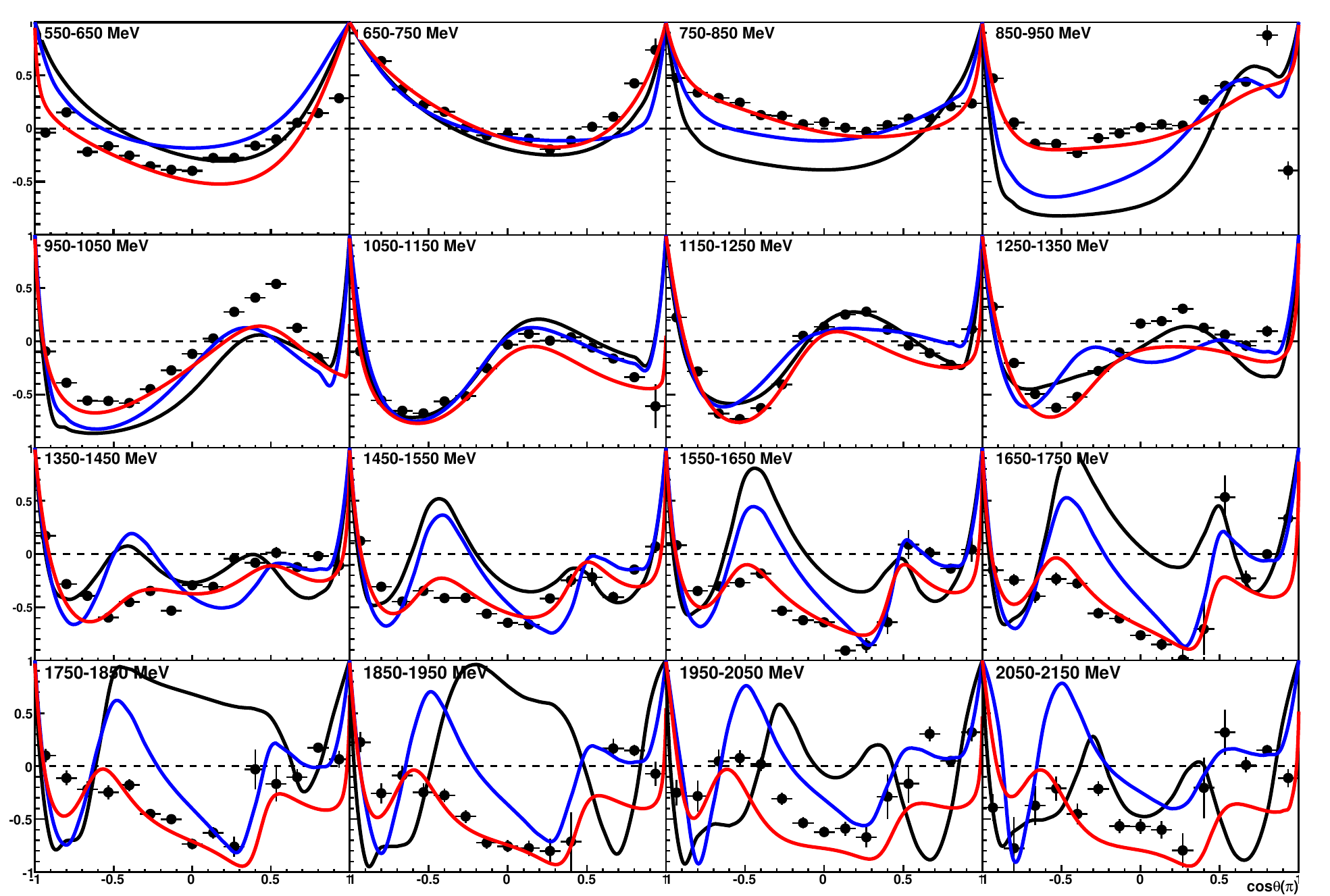}
      \put(17,18){\includegraphics[width=0.65\textwidth]{preliminary}}
      \put(0,65){\begin{scriptsize}E\end{scriptsize}}
    \end{overpic}
    \caption{Observable E for the reaction $\vec{\gamma}\vec{p} \to p \pi^0$ as a function of $\cos(\theta^\mathrm{CMS}_{\pi^0})$ compared to the predictions of the MAID (black), SAID (blue), and BnGa (red) analyses.}
    \label{fig:E_pion}
  \end{center}
\end{figure}

Preliminary results for the double polarisation observable E in the reaction $\vec{\gamma}\vec{p} \to p \pi^0$ are shown in Fig. \ref{fig:E_pion} as a function of $\cos(\theta^\mathrm{CMS}_{\pi^0})$. The shown error bars so far only include statistical uncertainties. Angular dependent structures appear up to the highest measured energies. 
While data and predictions agree still well at lower energies, the analyses contradict each other for example between the second and third resonance region ($E_{\gamma}=850-950~\mathrm{MeV}$). In the third resonance region ($E_{\gamma}=950-1050~\mathrm{MeV}$), none of the predictions is able to describe the data and strong deviations appear at higher energies where also the predictions show significant differences.

\section{Summary}
The preliminary results for the measurement of the double polarisation observable E in the reactions $\vec{\gamma}\vec{p} \to p \pi^0$ and $\vec{\gamma}\vec{p} \to p \eta$ with the Crystal Barrel/TAPS experiment look very promising.
Combined with the data obtained with a linearly polarised photon beam\cite{thiel:2010} and the transversely polarised target\cite{hartmann:2011} this measurement is an important step towards a complete experiment and 
will provide further constraints for the PWA.

\acknowledgements{%
This work was supported by the \emph{Deutsche Forschungsgemeinschaft} within SFB/TR-16.
}


%

}  


\end{document}